\documentclass{PoS}

\title{Baikal-GVD: first cluster Dubna}

\ShortTitle{Baikal-GVD}

\author{A.D. Avrorin$^{a}$, A.V. Avrorin$^{a}$, V.M. Aynutdinov$^{a}$, R. Bannash$^{g}$, I.A. Belolaptikov$^{b}$, D.Yu. Bogorodsky$^{c}$, 
	V.B. Brudanin$^{b}$, N.M. Budnev$^{c}$, I.A. Danilchenko$^{a}$, S.V. Demidov$^{a}$, G.V. Domogatsky$^{a}$, A.A. Doroshenko$^{a}$, A.N. Dyachok$^{c}$, 
	Zh.-A.M. Dzhilkibaev$^{a}$, S.V. Fialkovsky$^{e}$, A.R. Gafarov$^{c}$, O.N. Gaponenko$^{a}$, K.V. Golubkov$^{a}$, T.I. Gress$^{c}$, 
	Z. Honz$^{b}$, K.G. Kebkal$^{g}$, O.G. Kebkal$^{g}$, K.V. Konischev$^{b}$, A.V. Korobchenko$^{b}$, A.P. Koshechkin$^{a}$, F.K. Koshel$^{a}$, 
	A.V. Kozhin$^{d}$, V.F. Kulepov$^{e}$, D.A. Kuleshov$^{a}$, V.I. Ljashuk$^{a}$, M.B. Milenin$^{e}$, R.A. Mirgazov$^{c}$, E.R. Osipova$^{d}$, 
	A.I. Panfilov$^{a}$, L.V. Pan'kov$^{c}$, E.N. Pliskovsky$^{b}$, M.I. Rozanov$^{f}$, E.V. Rjabov$^{c}$, B.A. Shaybonov$^{b}$, 
	A.A. Sheifler$^{a}$, M.D. Shelepov$^{a}$, A.V. Skurihin$^{d}$, A.A. Smagina$^{b}$, \speaker{O.V. Suvorova}$^{a}$, V.A. Tabolenko$^{c}$, 
        B.A. Tarashansky$^{c}$, S.A. Yakovlev$^{g}$, A.V. Zagorodnikov$^{c}$, V.A. Zhukov$^{a}$, and V.L. Zurbanov$^{c}$
\\		
		$^a$Institute for Nuclear Research, Moscow, 117312 Russia\\
		$^b$Joint Institute for Nuclear Research, Dubna, 141980 Russia\\
		$^c$Irkutsk State University, Irkutsk, 664003 Russia\\		
		$^d$Institute of Nuclear Physics, Moscow State University, Moscow, 119991 Russia\\
		$^e$Nizhni Novgorod State Technical University, Nizhni Novgorod, 603950 Russia\\
		$^f$St. Petersburg State Marine Technical University, St. Petersburg, 190008 Russia\\
		$^g$EvoLogics, Germany
        \\
        E-mail: \email{suvorova@cpc.inr.ac.ru}}

\abstract{In April 2015 the demonstration cluster "Dubna" was deployed
and started to take data in Lake Baikal. This array is the first cluster of
the cubic kilometer scale Gigaton Volume Detector (Baikal-GVD), which is
constructed in Lake Baikal. In this contribution we will review the design
and status of the array.\
}

\FullConference{The European Physical Society Conference on High Energy Physics\\
		22--29 July 2015\\
		Vienna, Austria}

\begin{document}

\section{Introduction}
The neutrino telescope Baikal Gigaton Volume Detector (GVD) is currently under construction in Lake Baikal~\cite{b2},
which is a deep underwater Cherenkov detector of next generation. Main stream of neutrino
studies with Baikal-GVD is search for astrophysical neutrinos incoming from lower hemisphere 
with energies higher a few tens of TeV. 
Potential astrophysical targets for neutrino detection 
with GVD are celestial objects which are visible in GeV-TeV gamma-rays (SNRs, AGNs, GRBs and so on) 
or invisible dark matter in the halo of the Milky Way, the Galactic Center (GC), dwarfs, and also in
the center of the Sun and the Earth. At 51$^\circ$ North latitude of Baikal site visibility time
is more than 65\% of observation time for
directions on Southern sky from where a most part of first candidates on cosmic neutrinos has been detected 
by IceCube~\cite{IC2}. They have not been identified with known sources because of 10-15 degree 
uncertainties in reconstruction of neutrino induced cascades in polar ice. Unique  optical properties of
the Baikal deepwater give
an advantadge in angular resolution down to 4 degrees or less in detection of Cherenkov photons from 
electro-magnetic or hadron showers produced by high energy neutrinos in Lake Baikal. 

\begin{figure}[!htb]
\begin{center}
\begin{tabular}{cc}
\includegraphics[width=0.45\textwidth,angle=0]{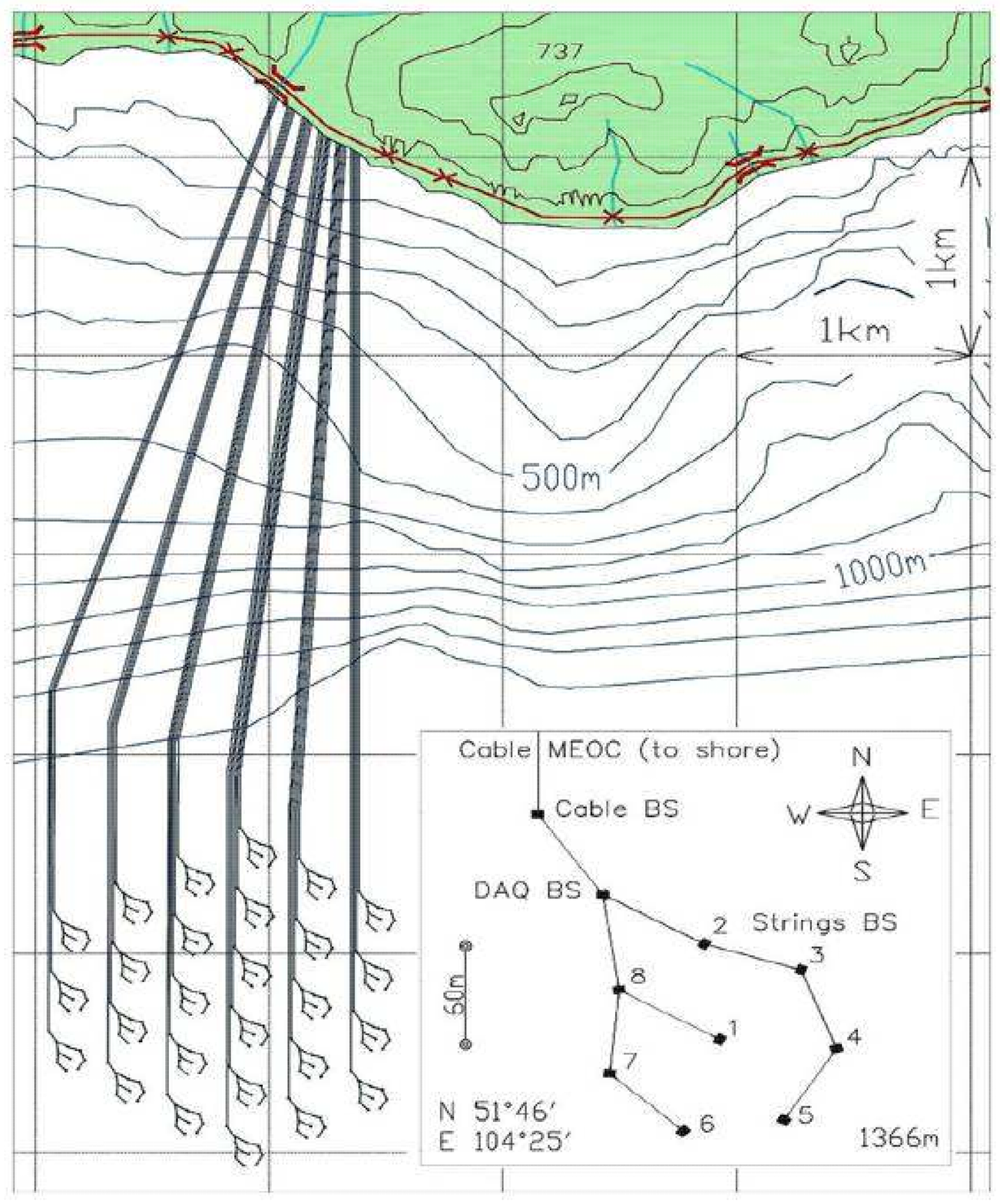} 
&
\includegraphics[width=0.5\textwidth,angle=0]{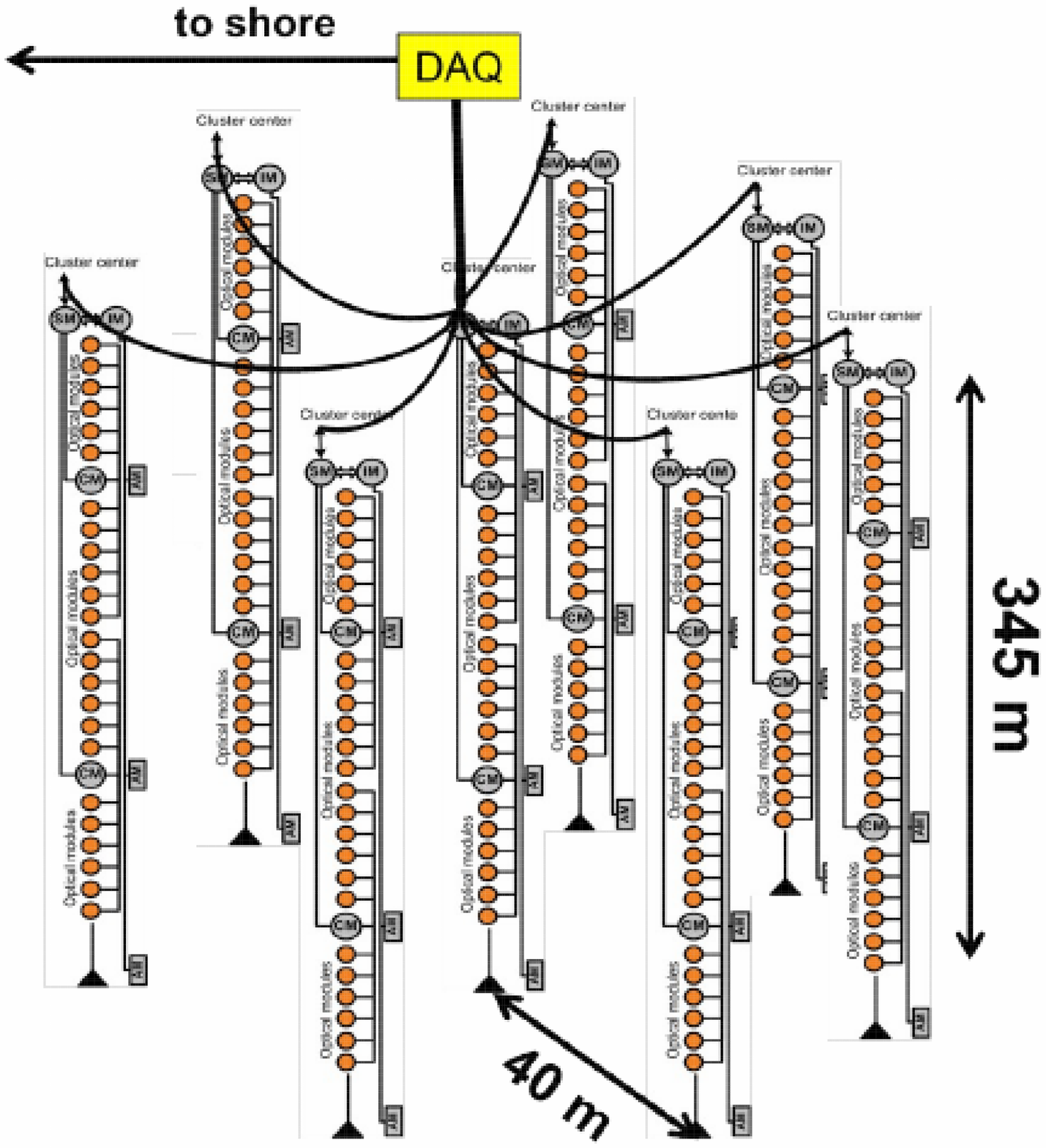} 
\end{tabular}
\end{center}
\caption{\label{fig:mapGVD} Artistic views. Left: Layout of the GVD. In inner box the one cluster is shown. 
Right: The first GVD cluster Dubna.} 
\end{figure}

The site chosen for the experiment is in the southern basin of Lake Baikal. Here, the combination of hydrological, 
hydro-physical, and landscape factors is studied to be optimal for deployment and operation of the neutrino telescope. 
The water depth is about 1360 m at distances beginning from about of three kilometers from the shore. 
The water transparency is characterized by an absorption length of about 20--25 m and a scattering length of 30--50 m~\cite{b2}. 
The water chemiluminescence is moderate at the detector site. Baikal-GVD is a three-dimensional lattice of optical modules (OMs), 
that is photomultiplier tubes housed in transparent pressure spheres, arranged at vertical load-carrying cables to form strings. 
The telescope has a modular structure and consists of clusters of strings, functionally independent sub-arrays, 
which are connected to shore by individual electro-optical cables as shown in Fig.~\ref{fig:mapGVD} (left). Each cluster 
has a central string identical to seven others distant at radius of 60 meters in baseline configuration.

\section{First demonstration cluster - Dubna}
In April 2015 the first Baikal-GVD cluster named $Dubna$
was deployed and started operation in Lake Baikal following timetable of the project~\cite{b2,b3}. The Dubna cluster 
encloses 1.7 Megatons of fresh Baikal water. The cluster comprises a total 
of 192 optical modules arranged at eight 345 m long strings, 
as well as an acoustic positioning system. There is also an instrumentation string with equipment for array calibration 
and monitoring of environmental parameters. Artistic view of the Dubna cluster is shown in Fig.~\ref{fig:mapGVD} (right).  
Each string comprises 24 OMs spaced by 15 m at depths of 900 m to 1250 m below the surface. 
In 2015 seven side strings have been located at a reduced radius of 40 m around a central one. The reason is to 
increase the sensitivity to low-energy  atmospheric muons and neutrinos which are used for array calibration. 
In the next year, strings will be moved to the baseline distances. 

A section is the basic detection unit of the GVD neutrino telescope. 
It comprises 12 optical modules (OM) and a central module (CeM). Among the basic functions of OMs 
are detection of the particle radiation; shaping of the output analog pulse for signal transmission to the ADC board;
control of the PMT operation modes; calibration and monitoring of the parameters of OM electronic components.
The block diagram of a section and OM view are published elsewere~\cite{b3,daq,OM}. Each optical module consists of a pressure-resistant 
glass sphere of 43.2 cm diameter which holds the OM electronics and the PMT which is surrounded by a high permittivity 
alloy cage for shielding it against the Earth magnetic field. A large photomultiplier tube Hamamatsu R7081-100 with a 10-inch 
hemispherical photocathode and quantum efficiency up to 35\% has been selected as light sensor. Besides the PMT, an OM comprises 
a high voltage power supply unit (HV), a fast two-channel preamplifier, and a controller. For time and amplitude calibration 
of the measuring channel, two LEDs are installed in the optical module. The OM controller is intended for HV control and monitoring 
for PMT noise measurements and for time and amplitude calibration~\cite{daq}. 

The PMT signals from all OMs are transmitted to the CeM via 90 meters of coaxial cables, where they 
are digitized by custom-made 12-channel ADC boards with 200 MHz sampling rate. 
The slow-control board located in the CeM provides data communication between OM and CeM via an underwater RS-485 bus. 
Also, this unit is intended for OM power control (to switch power on/off for each optical module independently). 
The ADC board provides trigger logic, data readout and digital processing, and connection via local Ethernet 
to the cluster DAQ center, control of the section operation and the section trigger logic. A request analyzer 
forms the section trigger request (local trigger) on the basis of channel requests L (low channel threshold, ~0.3 p.e.) 
and H (high threshold, ~3 p.e.) from 12 ADC channels. This unit contains a programmable coincidence matrix (12Hx12L), 
which provides a simple way to generate the section trigger request. There are two basic trigger modes: 
(A) coincidences of $>$N L-requests within a selectable time window, or (B) coincidences of L and H requests from 
any neighbouring OMs within a section. A request of the section trigger is transferred from the Master board through a string 
communication module (CoM) to the cluster DAQ-center, where a global trigger for all sections is generated. 
Data from the strings are transferred through DSL-modem Ethernet channel to the cluster center. The data transmission between 
the cluster DAQ-center and shore station is provided through optical fiber lines extended at about 6 km.
\begin{figure}[!htb]
\begin{center}
\begin{tabular}{cc}
\includegraphics[width=0.45\textwidth,angle=0]{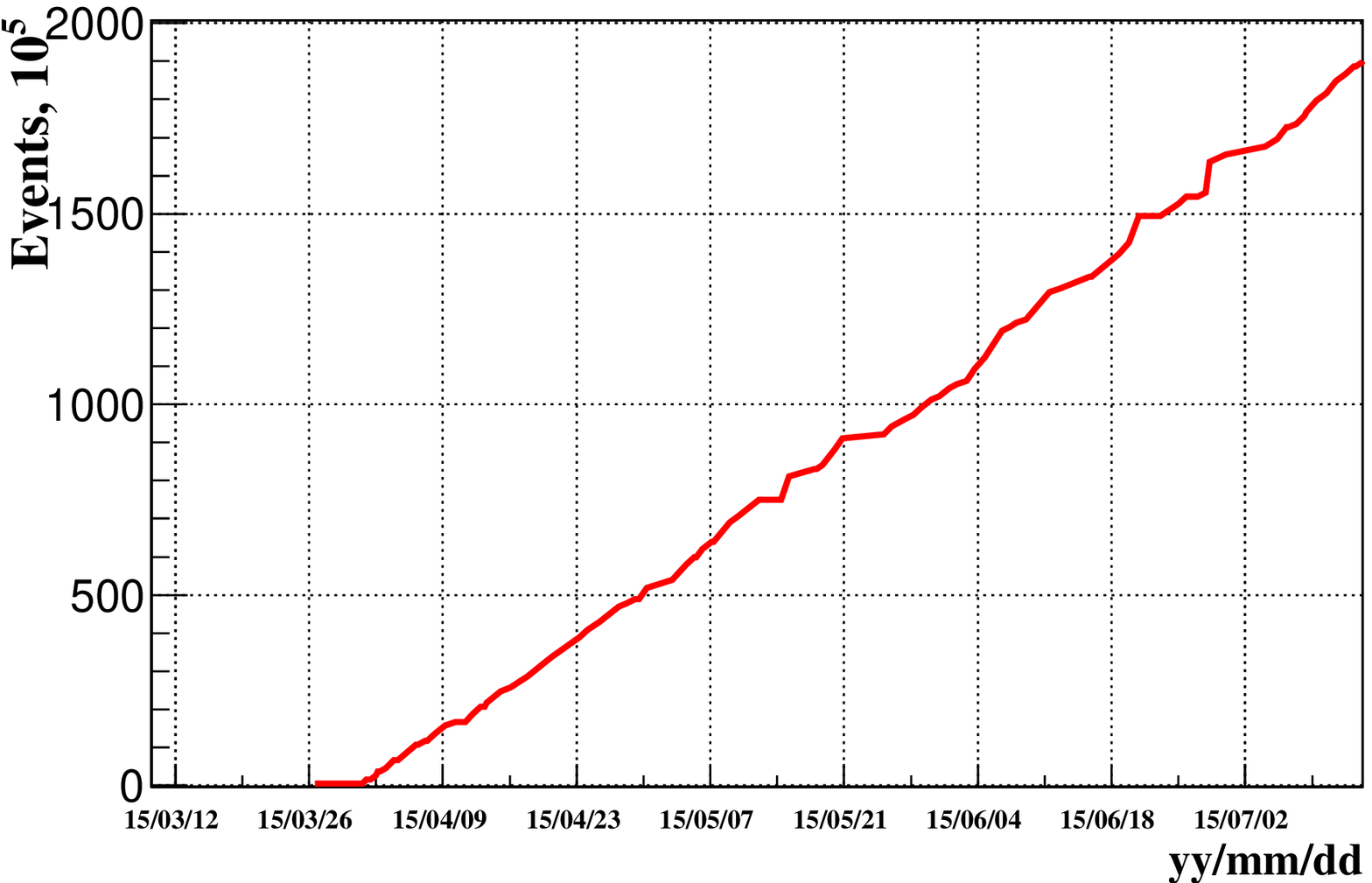} 
&
\includegraphics[width=0.5\textwidth,angle=0]{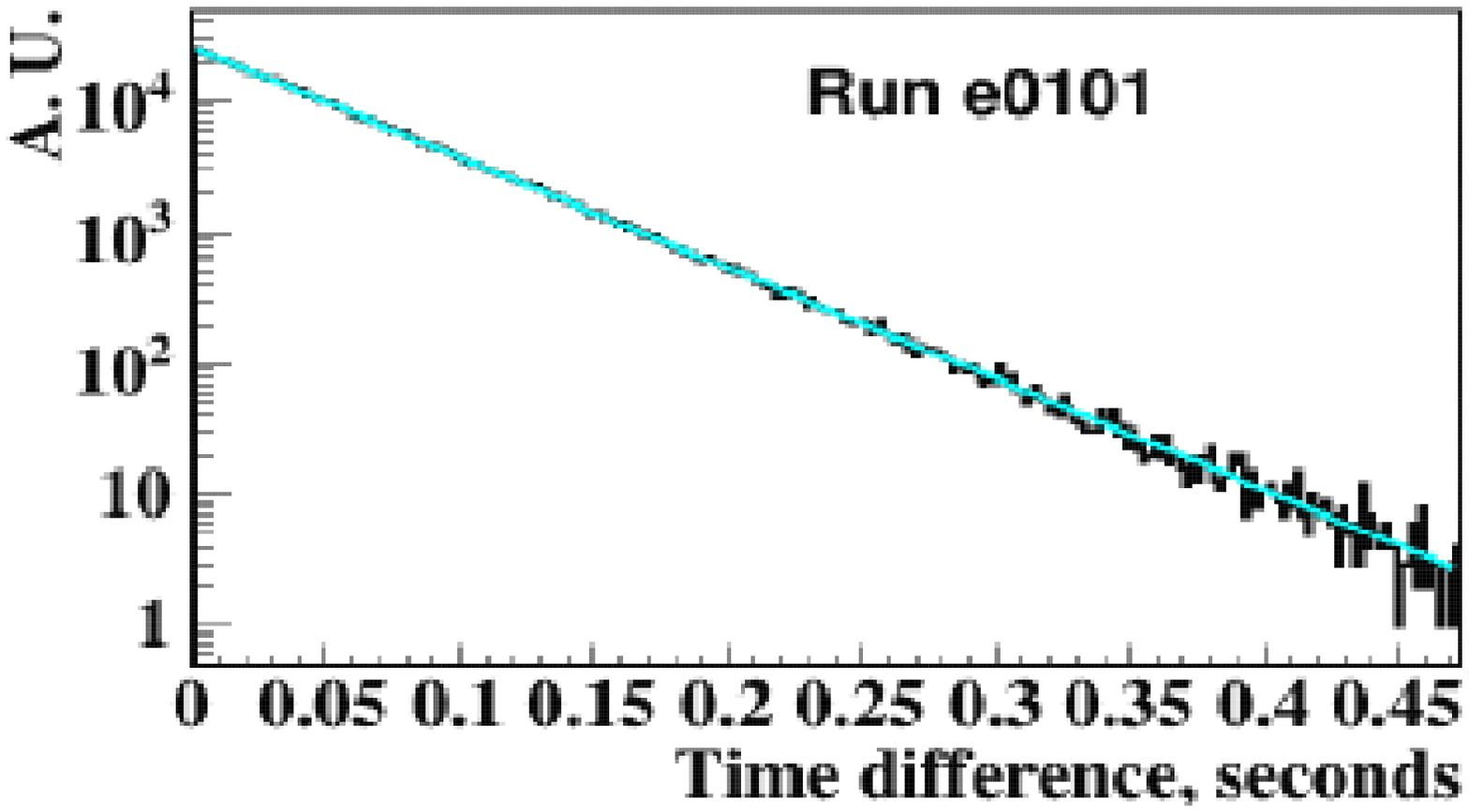} 
\end{tabular}
\end{center}
\caption{\label{fig:stabDubna} Left: Integrated number of recorded events since April 2015. 
Right: The time difference between subsequent events is shown for one run.
} 
\end{figure}

In 2015 the Dubna cluster is operating in several testing and data taking modes. 
Since Aprile till June about 1.7$\cdot$10$^8$ events have been recorded. Stability and efficiency of the cluster operation 
during 2015 are illustrated in Fig. \ref{fig:stabDubna} (left) by integral rate of general trigger. 
Quality of data is seen in Fig. \ref{fig:stabDubna} (right) with distribution of time difference between 
subsequent events for one run.  Obtained exponential behavior is consistent with expectation for randomly distributed 
experimental events. 

\section{Performance and sensitivity}
\begin{figure}[!htb]
\begin{center}
\begin{tabular}{cc}
\includegraphics[width=0.48\textwidth,angle=0]{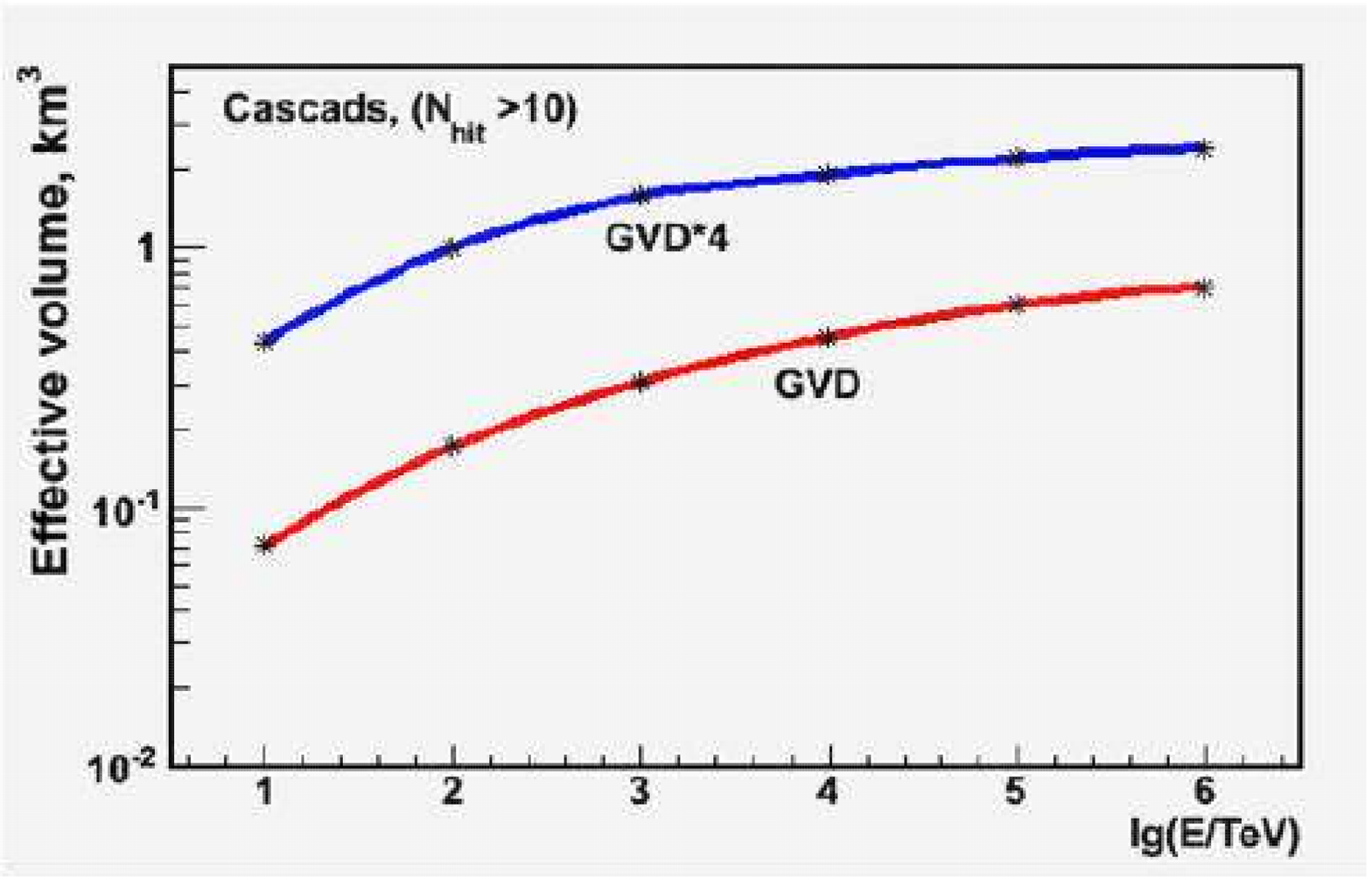} 
&
\includegraphics[width=0.48\textwidth,angle=0]{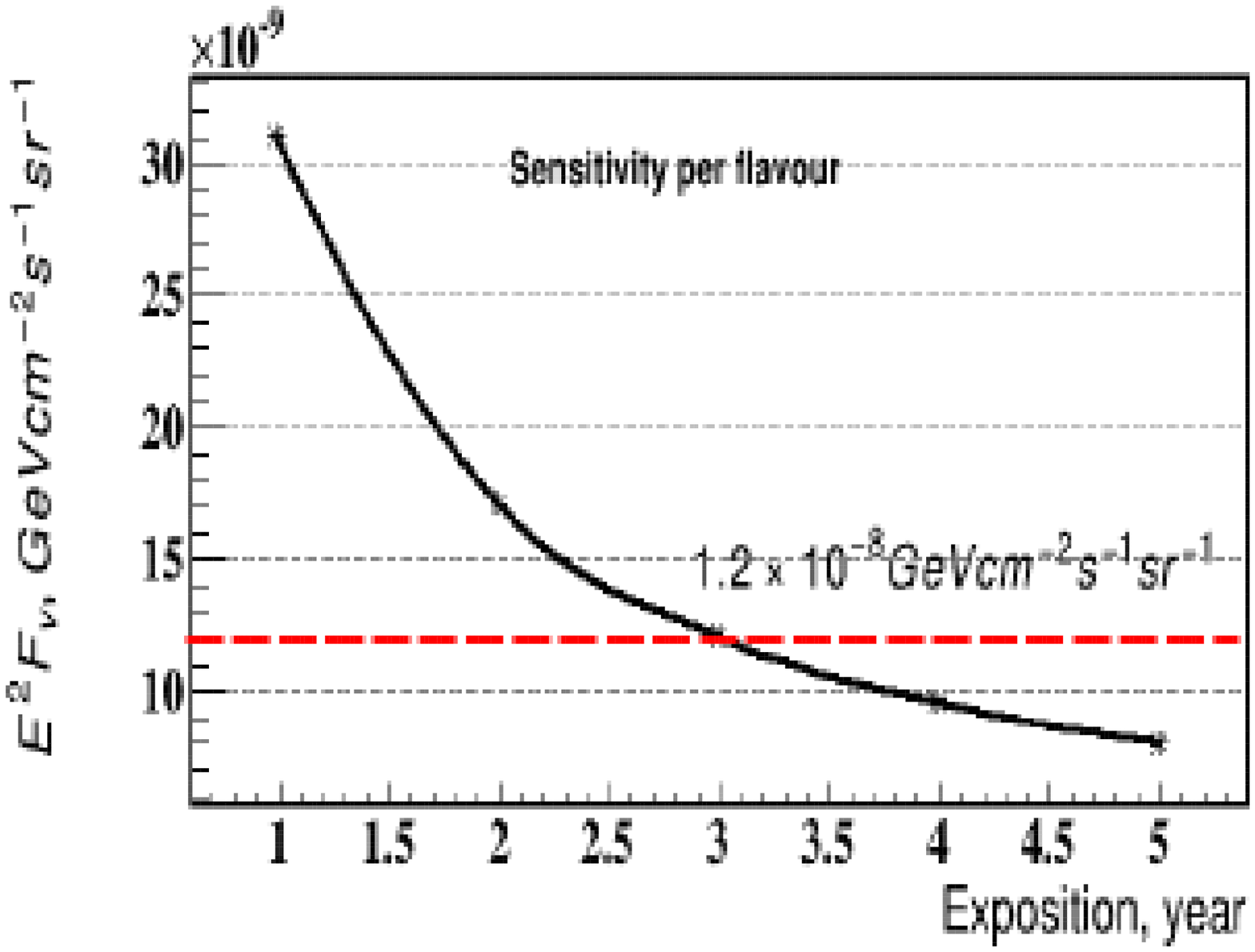} 
\end{tabular}
\end{center}
\caption{\label{fig:showsGVD} Sensitivities of GVD configurations to neutrino induced showers. 
Left: Effective volumes of cascades detection. The curves labeled by GVD$*$4 and GVD relate to 
configurations with 10368 OMs and 2304 OMs, respectively. Right: Cluster sensitivity for one 
flavor neutrino flux with an E$^{-2}$ spectrum as function of the observation years. 
The long-dashed line indicates the one flavor neutrino flux value obtained by IceCube.
}
\end{figure}

The first cluster of Baikal-GVD in its baseline configuration will have the potential to detect astrophysical 
neutrinos with a flux value measured recently by IceCube~\cite{IC2}. The search for high-energy neutrinos is based 
on the selection of cascade events generated by neutrino interactions in the sensitive volume of array. 
After applying an iterative procedure of vertex reconstruction followed by the rejection of hits contradicting 
the cascade hypothesis on each iteration stage, events with a final multiplicity of hit OMs $N_{hit}>$ 20 are selected 
as high-energy neutrino events. Shower effective volumes for two GVD configurations are shown in Fig.\ref{fig:showsGVD}
(left). Shower effective volumes (11/3 condition - at least 11 hit OMs on at least 3 strings) for GVD*4 are 
about of 0.4--2.4 km$^3$ above 10 TeV. The accuracy of shower energy reconstruction with GVD configuration of 10368 OMs
is about of 20--35\% depending
on shower energy, while directional resolution (median value) is 4$^{\circ}$, which is substantially better 
than the 10--15 degrees accuracy 
for IceCube~\cite{IC2}. The expected number of background
events from atmospheric neutrinos is strongly suppressed for energies higher than 100 TeV. We
expect about one event per year with $E_{sh}>$100 TeV from an all-flavor astrophysical flux in GVD-cluster
with the normalization E$^2\times$ Flux = 3.6$\cdot$10$^{-8}$~GeV~cm$^{-2}$~s$^{-1}$~sr$^{-1}$, compared 
to about 10 events in IceCube. Preliminary estimate of the cluster sensitivity to one flavor neutrino flux with an $E^{-2}$ 
spectrum and flavor ratio 1:1:1 for all--flavor flux as function of the observation years is shown in Fig.\ref{fig:showsGVD} (right),
with no systematics accounts. Three year exposition allows sensitivity at a level of flux value measured by IceCube.
\begin{figure}[!htb]
\begin{center}
\begin{tabular}{cc}
\includegraphics[width=0.34\textwidth,angle=-90]{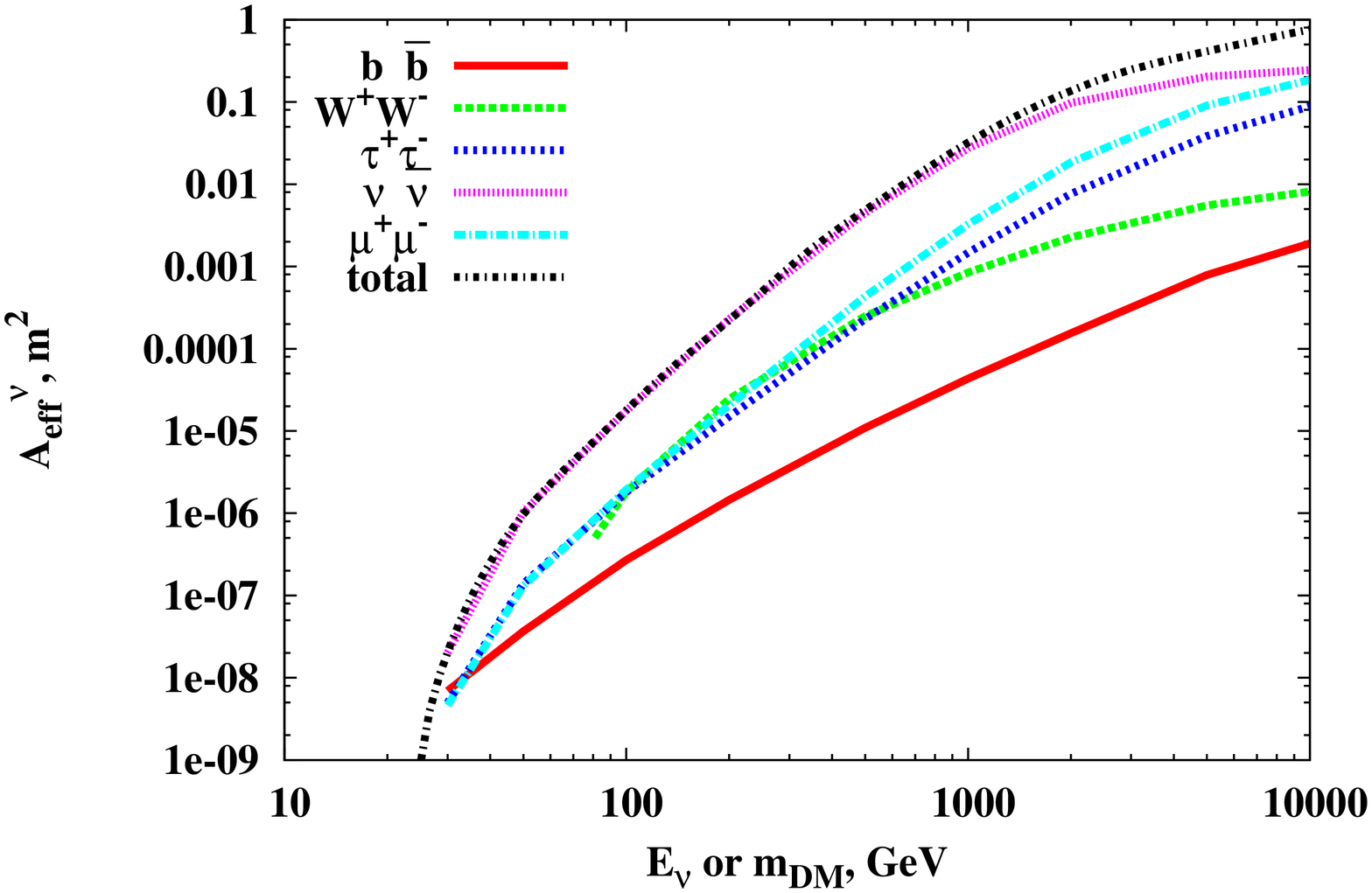} 
&
\includegraphics[width=0.34\textwidth,angle=-90]{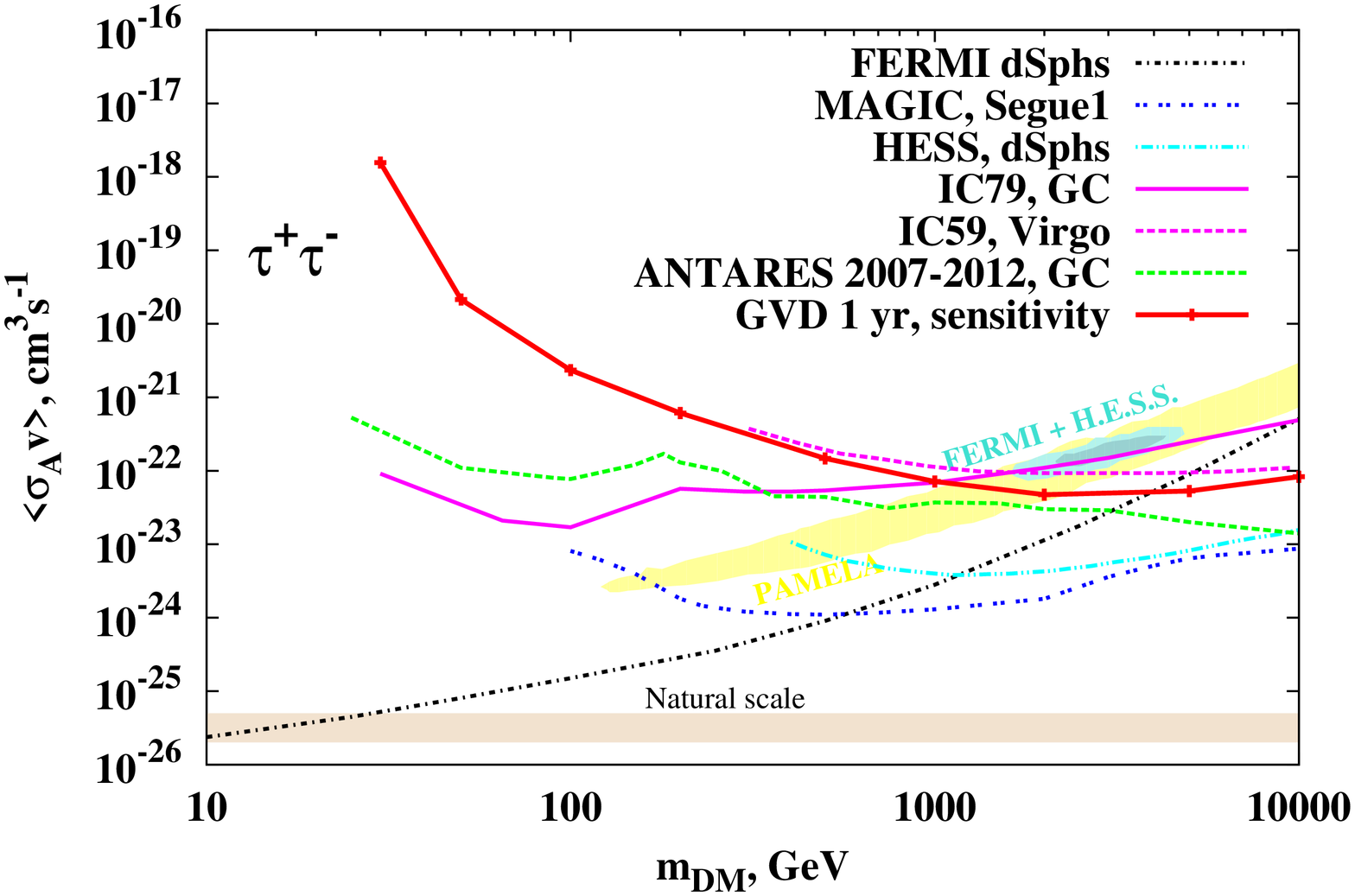} 
\end{tabular}
\end{center}
\caption{\label{fig:DMtauGC} Left: Neutrino effective area of single GVD cluster 
(total, black) and averaged over neutrino spectra effective areas for different channels (color).
Right: Sensitivity of GVD to tau-channel of dark matter annihilations in the Galactic Center 
in comparison with other experiments.}
\end{figure}

We studied the GVD sensitivity with 12 clusters to neutrinos from dark matter annihilations in the Galactic Center 
for 1 year of livetime observation.\footnote{For case of DM decays in the GC and for more details see Ref.~\cite{DM} and references therein.} 
For this study we apply a muon trigger 
formed by requirements to select events with at least 6 fired OMs on at least 3 strings (3/6).  The neutrino 
effective area for selection (3/6) as a function of neutrino energy for one cluster is presented in Fig.\ref{fig:DMtauGC} (left)
by black line. We choose a search region as a cone around the direction towards the GC with half angle $\psi_0$. 
The expected number of signal events in the search region for the livetime $T$ is
estimated as follow 
\begin{equation}
\label{eq:6}
N(\psi_0) = T\frac{\langle\sigma_A v\rangle R_0\rho_{local}^{2}}
{8{\pi} m_{DM}^2}J_{\Delta \Omega}\int
dE \cdot S(E)\frac{dN_{\nu}}{dE}.
\end{equation}
Here $S(E)$ is neutrino effective area of the telescope. In Fig.~\ref{fig:DMtauGC} (left) along with effective area of one 
cluster (black) are shown effective areas averaged over neutrino spectrum $\frac{dN_{\nu}}{dE}$ in given annihilation channel. 

We consider $b\bar{b}$, $\tau^{+}\tau^{-}$, $\mu^{+}\mu^{-}$, $W^{+}W^{-}$ and $\nu\bar{\nu}$ channels, where in the 
latter case we assume flavor symmetric annihilation. Neutrino spectra from dark matter annihilation have been taken 
from~\cite{Baratella:2013fya}. The astrophysical factor $J_{\Delta \Omega}$ here is an average value over the search region.
The expected upper bounds on dark matter annihilation cross section have been obtained from this equation.
There are several theoretical uncertainties in the number of signal events related to neutrino oscillation parameters, 
neutrino-nucleon cross section etc. However, the most important of them is the uncertainty related to lack 
of knowledge of dark matter density profile near the GC. Our study shows that 1 year GVD sensitivity with incorporated 
realistic efficiency and systematic uncertainties achieves values $5\cdot 10^{-24}$~cm$^3$s$^{-1}$ for dark matter annihilation 
cross section and $2.4\cdot 10^{26}$~s for DM lifetime in the most energetic $\nu\bar{\nu}$ channel.

To summarize, since April 2015 the data taking with the first full-completed cluster Dubna of the 
Baikal Gigaton Volume Detector has been started. The array comprises 192 optical modules. The modules 
are arranged at depths down to 1,300 m. Over its next stages of construction, the telescope 
will be stepwise extended by deploying new clusters. By 2020, it is planned to be consisted of 10-12 
clusters with a total volume of about 0.4 cubic kilometers. 

The work of S.V.~Demidov and O.V.~Suvorova was supported by the RSCF
grant 14-12-01430.

\end{document}